\documentclass[a4paper,11pt]{article}
\usepackage[dvips]{color,graphicx}
\usepackage{stmaryrd}
\def\dprod{\displaystyle\prod}

\def\dsum{\displaystyle\sum}

\usepackage{graphicx,amssymb,amsmath,bm,latexsym}

\allowdisplaybreaks

\begin{document}
%\begin{titlepage}
\begin{center}
{\Large\bf Exact correlators in the Gaussian Hermitian matrix model}\vskip .2in
{\large Bei Kang$^{a}$,
Ke Wu$^{a}$,
Zhao-Wen Yan$^{b}$,
Jie Yang$^{a}$,
Wei-Zhong Zhao$^{a}$\footnote{Corresponding author: zhaowz@cnu.edu.cn}} \vskip .2in
$^a${\em School of Mathematical Sciences, Capital Normal University,
Beijing 100048, China} \\
$^b${\em School of Mathematical Sciences, Inner Mongolia University, Hohhot
010021, China}\\

\begin{abstract}
We present the $W_{1+\infty}$ constraints for the Gaussian Hermitian matrix model, where
the constructed constraint operators yield the $W_{1+\infty}$ $n$-algebra.
For the Virasoro constraints, we note that the constraint operators give the null 3-algebra.
With the help of our Virasoro constraints, we derive a new effective formula for correlators
in the Gaussian Hermitian matrix model.

\end{abstract}

\end{center}

{\small Keywords: Conformal and $W$ Symmetry, Matrix Models, $n$-algebra}
%\vfill

%\end{titlepage}

\section{Introduction}
The various constraints for matrix models have attracted remarkable attention,
such as Virasoro/$W$-constraints \cite{Mironov}-\cite{Itoyama} and
Ding-Iohara-Miki constraints \cite{Zenkevich, ZenkevichJHEP}.
Due to the Bagger-Lambert-Gustavsson (BLG) theory of
M2-branes \cite{BL2007, Gustavsson}, $n$-algebra and its applications
have aroused much interest \cite{Izquierdo}-\cite{Zhang}.
In the context of matrix models, usually the Virasoro/W-constraint operators
do not yield the closed $n$-algebra.
Whether there exist such kind constraint operators leading to the closed $n$-algebra
has recently been investigated for the (elliptic) Hermitian one-matrix models.
By inserting the special multi-variable realizations of
the $W_{1+\infty}$ algebra under the integral, it was found that the derived
constraint operators for the Hermitian one-matrix model may yield the closed
$W_{1+\infty}$ ($n$-)algebras \cite{Wangr1}.
For the case of the elliptic matrix model, one can obtain the constraint operators
associating with the $q$-operators \cite{Nedelin, Wangr2}.
The situation is different from that of the Hermitian one-matrix model,
since the derived constraint operators do not yield the closed algebra.
However, it was shown that the ($n$-)commutators of the constraint operators
are compatible with the desired generalized $q$-$W_{\infty}$ ($n$-)algebras
once we act on the partition function \cite{Wangr2}.

The partition functions of various matrix models can be obtained by
acting on elementary functions with exponents of the given operators.
For the Gaussian Hermitian matrix model, its partition function is generated
by the operator $\hat W_{-2}$ \cite{Shakirov2009}.
This operator is also the constraint operator for the Hermitian one-matrix model which
associates with the Lassalle operator and the potential of
the $A_{N-1}$-Calogero model \cite{Wangr1}.
The correlators in the Gaussian Hermitian matrix model have been well investigated
\cite{Harer}-\cite{1706.03667}. A compact formula for correlators  has been given
by finite sums over Young diagrams of a given size, which involve also the well known
characters of symmetric group \cite{1705.00976}.
Moreover, the $2m$-fold Gaussian correlators of rank $r$ tensors have been given
by $r$-linear combinations of dimensions with the Young diagrams of size $m$ \cite{1706.03667}.
In this letter, we reinvestigate the Gaussian Hermitian matrix model
and present its Virasoro/$W$-constraints. We intend to further explore the properties
of the constraints and derive a new formula for correlators in this matrix model.

%%%%%%%%%%%%%%%%%%%%%%%%%%%%%%%%%%%%%%%%%%%%%%%%%%%%%%%%%%%%%%%%%%%%%%%%%%%%%%%%%%%%%%%%%
\section{$W_{1+\infty}$ constraints for the Gaussian Hermitian matrix model}
%%%%%%%%%%%%%%%%%%%%%%%%%%%%%%%%%%%%%%%%%%%%%%%%%%%%%%%%%%%%%%%%%%%%%%%%%%%%%%%%%%%%%%%%%
Let us consider the Gaussian Hermitian matrix model
\begin{eqnarray}\label{GHMM}
Z_{G}&=&\int_{N\times N}d\phi exp({-tr\phi^{2}/2+\sum_{k= 0}^{\infty}t_{k}tr\phi^{k}})\nonumber\\
&=&e^{Nt_{0}}(1+C_{i_1}(N)t_{i_1}+\frac{1}{2!}C_{i_1i_2}(N)t_{i_1}t_{i_2}+\frac{1}{3!}C_{i_1i_2i_3}(N)t_{i
_1}t_{i_2}t_{i_3}+\cdots),
\end{eqnarray}
where the coefficients $C_{i_{1}\cdots i_{l}}(N)$ are the so-called $l$-point correlators,
which are given by the Gaussian integrals
\begin{eqnarray}
C_{i_{1}\cdots i_{l}}(N)=\langle tr\phi^{i_{1}}\cdots tr\phi^{i_{l}}\rangle
=\int_{N\times N}d\phi tr\phi^{i_{1}}\cdots tr\phi^{i_{l}}exp(-\frac{1}{2}tr\phi^{2}).
\end{eqnarray}
Due to the reflection symmetry of the action $tr\phi^{2}$, when $i_{1}+\cdots + i_{l}$ is odd,
we have $C_{i_{1}\cdots i_{l}}(N)=0$.

The partition function of the Gaussian model (\ref{GHMM}) can also be expressed as \cite{Shakirov2009}
\begin{eqnarray}
Z_{G}=\sum_{s=0}^{\infty}Z_G^{(s)}
=e^{\hat{W}_{-2}/2}e^{Nt_0},
\end{eqnarray}
where
\begin{eqnarray}
Z_G^{(s)}=e^{Nt_0}\sum_{l=0}^{\infty}\sum_{\substack{i_1+\cdots +i_l=s\\i_1,\cdots,i_l\geqslant 1}}
\langle tr\phi^{i_1}\cdots tr\phi^{i_l}\rangle \frac{t_{i_1}\cdots t_{i_l}}{l!},
\end{eqnarray}
and the operator $\hat{W}_{-2}$ is given by
\begin{eqnarray}
\hat{W}_{-2}=\sum_{j_1,j_2=0}^{\infty}(j_1j_2t_{j_1}t_{j_2}\frac{\partial}{\partial
t_{j_1+j_2-2}}+(j_1+j_2+2)t_{j_1+j_2+2}\frac{\partial}{\partial t_{j_1}}\frac{\partial}{\partial t_{j_2}}).
\end{eqnarray}
It indicates that the partition function (\ref{GHMM}) can indeed be generated by the operator $\hat{W}_{-2}$.

The action of the operator $\hat{W}_{-2}$ on $Z_G^{(s)}$ leads to increase
the grading in the following sense:
\begin{eqnarray}\label{increa}
\hat{W}_{-2}Z_G^{(s)}=(s+2)Z_G^{(s+2)}.
\end{eqnarray}
The operator preserving the grading is given by \cite{Shakirov2009}
\begin{eqnarray}
\hat{D}=\sum_{j=0}^{\infty}jt_{j}\frac{\partial}{\partial t_{j}},
\end{eqnarray}
which acting on $Z_G^{(s)}$ gives
\begin{eqnarray}\label{dzs}
\hat D Z_G^{(s)}=sZ_G^{(s)}.
\end{eqnarray}
The commutation relation between $\hat{D}$ and $\hat{W}_{-2}$ is
\begin{eqnarray}\label{crdw}
[\hat{D}, \hat{W}_{-2}]=2\hat{W}_{-2}.
\end{eqnarray}
Note that the actions of $\hat D$ and $\hat{W}_{-2}$ on $Z_G$ give
\begin{eqnarray}\label{dwgr}
\hat D Z_G=\hat{W}_{-2}Z_G.
\end{eqnarray}

For the operators $\frac{\partial}{\partial t_{2}}$ and $\hat{D}$,
there is the similar commutation relation as (\ref{crdw})
\begin{eqnarray}\label{crtd}
[\hat{D}, \frac{\partial}{\partial t_{2}}]=-2\frac{\partial}{\partial t_{2}}.
\end{eqnarray}
The actions of $\frac{\partial}{\partial t_{2}}$ and $\hat D$ on $Z_G$ give
\begin{eqnarray}\label{dl0}
\frac{\partial}{\partial t_{2}}Z_{G}=(\hat D+N^{2})Z_{G}.
\end{eqnarray}
By means of (\ref{crtd}) and (\ref{dl0}), it is easy to show that
\begin{eqnarray}\label{decre}
\frac{\partial}{\partial t_{2}}Z_G^{(s)}=(s-2+N^2)Z_G^{(s-2)}.
\end{eqnarray}
In contrast with the operator $\hat{W}_{-2}$,
we see that the operator $\frac{\partial}{\partial t_{2}}$
decreases the grading in the sense (\ref{decre}).

Let us introduce the operators
\begin{eqnarray}\label{wmcope}
W_{m}^{r}=(-\frac{1}{2})^{r-1}(\hat{W}_{-2})^{m}
(\hat{W}_{-2}-\hat{D})^{r-1},
\, m,r\in \mathbb{N},\,r\geqslant2,
\end{eqnarray}
which obviously satisfy
\begin{equation}\label{wcon}
W_m^{r}Z_{G}=0.
\end{equation}

The remarkable property is that
these constraint operators yield
\begin{eqnarray}\label{Walgebra}
[W_{m_{1}}^{r_{1}},W_{m_{2}}^{r_{2}}]=(\sum_{k=0}^{r_1-1}C_{r_1-1}^{k}m_2^k
-\sum_{k=0}^{r_2-1}C_{r_2-1}^{k}m_1^k)W_{m_1+m_2}^{r_1+r_2-1-k},
\end{eqnarray}
and $n$-algebra
\begin{eqnarray}\label{NWnalgebra}
[W_{m_1}^{r_1}, W_{m_2}^{r_2}, \ldots,  W_{m_n}^{r_n}]
&:=&\epsilon_{1\ 2 \cdots n}^{i_1 i_2 \cdots i_n}
W_{m_{i_1}}^{r_{i_1}}W_{m_{i_2}}^{r_{i_2}}\cdots W_{m_{i_n}}^{r_{i_n}}\nonumber\\
&=&\epsilon_{1 \ 2 \cdots n}^{i_1 i_2 \cdots i_n} \dsum_{\alpha_1=0}^{\beta_1}
\dsum_{\alpha_2=0}^{\beta_2}
\cdots \dsum_{\alpha_{n-1}=0}^{\beta_{n-1}}
C_{\beta_1}^{\alpha_1}C_{\beta_2}^{\alpha_2}\cdots C_{\beta_{n-1}}^{\alpha_{n-1}}
\nonumber\\
&&\cdot m_{i_2}^{\alpha_1}m_{i_3}^{\alpha_2}\cdots m_{i_n}^{\alpha_{n-1}}
 W_{m_1+\cdots +m_{n}}^{r_1+\cdots+r_n-(n-1)-\alpha_1-\cdots-\alpha_{n-1}},
\end{eqnarray}
where $C_r^{k}=\frac{r(r-1)\cdots(r-k+1)}{k !}$,
$
\beta_{k}=\left\{
\begin{aligned}
&r_{i_{1}}-1 , & k= 1, \\
 & \sum_{j=1}^{k}r_{i_{j}}-k-\sum_{i=1}^{k-1}\alpha_{i}, & 2\leqslant k\leqslant n-1, \\
\end{aligned}
\right.
$
 and $\epsilon_{1 \ 2 \cdots n}^{i_1 i_2 \cdots i_n}$ is given by
$
\epsilon _{j_{1}\cdots j_{p}}^{i_{1}\cdots i_{p}}=\det \left(
\begin{array}{ccc}
\delta _{j_{1}}^{i_{1}} & \cdots & \delta _{j_{p}}^{i_{1}} \\
\vdots &  & \vdots \\
\delta _{j_{1}}^{i_{p}} & \cdots & \delta _{j_{p}}^{i_{p}}%
\end{array}%
\right).
$

It is noted that (\ref{Walgebra}) and (\ref{NWnalgebra}) completely match with the $W_{1+\infty}$
($n$-)algebras presented in Ref.\cite{Wangr1}.
The $W_{1+\infty}$ $n$-algebra (\ref{NWnalgebra}) with $n$ even
is a generalized Lie algebra (or higher order Lie algebra), which satisfies
the generalized Jacobi identity
\begin{equation}\label{GJI}
\epsilon _{1\  2\cdots (2n-1)}^{i_{1} i_2 \cdots i_{2n-1}}
[[A_{i_{1}},A_{i_{2}},\cdots ,A_{i_{n}}],A_{i_{n+1}},\cdots ,A_{i_{2n-1}}]=0.
\end{equation}
For the constraint operators $W_{m}^{r}$ (\ref{wmcope}) with fixed $r=n+1$,
by taking the appropriate scaling transformations, it is not difficult to show that
these operators constitute the subalgebras
\begin{eqnarray}\label{sub2nalgebra}
[W_{m_1}^{n+1}, W_{m_2}^{n+1}, \ldots , W_{m_{2n}}^{n+1}]=\dprod_{1\leqslant j<k\leqslant 2n}(m_k-m_j)
W_{m_1+\cdots +m_{2n}}^{n+1},
\end{eqnarray}
and
\begin{eqnarray}\label{nullsubalg}
[W_{m_1}^{n+1}, \ldots ,W_{m_{2n+1}}^{n+1}]=0.
\end{eqnarray}

For the $W_{1+\infty}$ constraints (\ref{wcon}),
it is noted that the constraint operators (\ref{wmcope})
contain the operators increasing and preserving the grading.
Let us now introduce the following operators in terms of the operators decreasing and preserving the grading:
\begin{eqnarray}\label{nvco}
\tilde{W}_{m}^{r}=(-\frac{1}{2})^{r-1}(\frac{\partial}{\partial t_{2}})^{m}
(\hat{D}-\frac{\partial}{\partial t_{2}}+N^2)^{r-1},\, m,r\in \mathbb{N},\,r\geqslant2.
\end{eqnarray}
Straightforward calculation shows that they also yield the $W_{1+\infty}$ algebra (\ref{Walgebra})
and $n$-algebra (\ref{NWnalgebra}).
By carrying out the action of the operators (\ref{nvco})
on the partition function of the Gaussian model, it gives another $W_{1+\infty}$ constraints
\begin{eqnarray}\label{nvcowc}
\tilde{W}_{m}^{r}Z_{G}=0.
\end{eqnarray}

\section{Correlators in the Gaussian Hermitian matrix model}
Let us first recall the correlators in the Gaussian Hermitian matrix model.
Harer and Zagier presented a generating function for exact (all-genera) 1-point correlators
in the Gaussian Hermitian matrix model \cite{Harer, Zuber},
\begin{eqnarray}\label{HZC}
\frac{C_{i}}{(i-1)!!}=\text{coefficient of } x^{i}\lambda^N \text{ in } \frac{\lambda}{1-\lambda}
\frac{1}{(1-\lambda)-(1+\lambda)x^2},
\end{eqnarray}
where $i$ is even.
By using Toda integrability of the model, Morozov and Shakirov derived
the 2-point generalization of the Harer-Zagier 1-point function \cite{Shakirov0906},
\begin{eqnarray}\label{MSC}
&&\frac{C_{(2k+1)(2m+1)}}{(2k+1)!!(2m+1)!!}=\text{coefficient of } x^{2k+1}y^{2m+1}\lambda^N \text{ in }
\frac{\lambda}{(\lambda-1)^{3/2}}\frac{arctan(\frac{xy\sqrt{\lambda-1}}{\sqrt{\lambda-1+(\lambda+1)
(x^2+y^2)}})}{\sqrt{\lambda-1+(\lambda+1)(x^2+y^2)}},\nonumber\\
&&\frac{C_{(2k)(2m)}}{(2k-1)!!(2m-1)!!}=\text{coefficient of } x^{2k}y^{2m}\lambda^N \text{ in }\nonumber\\
&&\frac{\lambda(\lambda+1)x^2y^2}{(1-\lambda)}(\lambda-1+(1+\lambda)(x^2+y^2))^{-1}
(\lambda-1+(1+\lambda)(x^2+y^2)+(\lambda-1)x^2y^2)^{-1}\nonumber\\
&&-\frac{\lambda(\lambda+1)xy}{(1-\lambda)^{3/2}}(\lambda-1+(\lambda+1)
(x^2+y^2))^{-3/2}arctan(\frac{xy\sqrt{\lambda-1}}{\sqrt{\lambda-1+(\lambda+1)
(x^2+y^2)}}).
\end{eqnarray}
However, it should be noted that it is difficult to give the higher correlators in
this way. Recently Mironov and Morozov presented a compact formula for correlators
by finite sums over Young diagrams of a given size \cite{1705.00976},
\begin{eqnarray}\label{MMC1}
C_{i_1i_2\cdots i_l}(N)\equiv\mathcal{O}_\Lambda=\sum_{R\vdash |\Lambda|} \frac{1}{d_R}
\chi_R\{t_n=\frac{1}{2}\delta_{n,2}\}\cdot D_R(N)\cdot \psi_R(\Lambda),
\end{eqnarray}
where $\Lambda=\{i_1\geqslant i_2 \geqslant \cdots \geqslant i_l>0\}$ and $R$ are the Young diagrams of
the given size $\sum_ki_k$, and $D_R(N)$, $\chi_R\{t\}$, $\psi_R(\Lambda)$ and $d_R$ are respectively
the dimension of representation $R$ for the linear group $GL(N)$, the linear character (Schur polynomial),
the symmetric group character and the dimension of representation $R$ of the symmetric group $S_{|R|}$ divided by $|R|!$.
Furthermore, a representation of the correlators in terms of permutations is given by  \cite{1706.03667}
\begin{eqnarray}\label{MMC2}
\mathcal{O}_\sigma=\sum_{R\vdash m} \varphi_R([2^m])\cdot D_R(N)\cdot \psi_R(\sigma),
\end{eqnarray}
where $\varphi_R([2^m])$ are the symmetric group characters.

Let us turn to consider the Virasoro constraints in (\ref{nvcowc})
\begin{eqnarray}\label{nvcm2}
\tilde{W}_{l}^{2}Z_{G}=0,\,\, l\in \mathbb{N}.
\end{eqnarray}
The constraint operators yield the Witt algebra
\begin{eqnarray}\label{witta}
[\tilde{W}_{l_{1}}^{2}, \tilde{W}_{l_{2}}^{2}]
=(l_{2}-l_{1})\tilde{W}_{l_{1}+l_{2}}^{2},
\end{eqnarray}
and null $3$-algebra
\begin{eqnarray}\label{n3witt}
[\tilde{W}_{l_{1}}^{2}, \tilde{W}_{l_{2}}^{2}, \tilde{W}_{l_{3}}^{2}]=0.
\end{eqnarray}

When $l\neq 0$, by using the expression (\ref{nvco}) to calculate left-hand side of (\ref{nvcm2}),
we obtain
\begin{eqnarray}\label{wm3}
&&\sum_{i=1}^{\infty}it_{i}(C_{\underbrace{2\cdots 2}_{l}\displaystyle i}(N)+\sum_{i_1=1}^{\infty}C_{\underbrace{2\cdots 2}_{l}\displaystyle ii_1}(N)
t_{i_1}+\frac{1}{2!}\sum_{i_1,i_2=1}^{\infty}C_{\underbrace{2\cdots 2}_{l}\displaystyle ii_1i_2}(N)t_{i_1}t_{i_2}+\cdots)\nonumber\\
&&+(N^{2}+2l)(C_{\underbrace{2\cdots 2}_{l}}(N)+\sum_{i_1=1}^{\infty}C_{\underbrace{2\cdots 2}_{l}\displaystyle i_1}(N)
t_{i_1}+\frac{1}{2!}\sum_{i_1,i_2=1}^{\infty}C_{\underbrace{2\cdots 2}_{l}\displaystyle i_1i_2}(N)t_{i_1}t_{i_2}+\cdots)\nonumber\\
&&-(C_{\underbrace{2\cdots 2}_{l+1}}(N)+\sum_{i_1=1}^{\infty}C_{\underbrace{2\cdots 2}_{l+1}\displaystyle i_1}(N)t_{i_1}+\frac{1}{2!}
\sum_{i_1,i_2=1}^{\infty}C_{\underbrace{2\cdots 2}_{l+1}\displaystyle i_1i_2}(N)t_{i_1}t_{i_2}+\cdots)=0.
\end{eqnarray}
From the fact that the constant term in the left-hand side
of (\ref{wm3}) should be zero, we have
\begin{eqnarray}\label{c2mr}
C_{\underbrace{2\cdots 2}_{l+1}}(N)=(N^{2}+2l)C_{\underbrace{2\cdots 2}_{l}}(N).
\end{eqnarray}
Taking the special constraint operator $\tilde{W}^2_0$ in (\ref{nvcm2}), it is easy to obtain
\begin{eqnarray}
C_{2}(N)=N^{2}.
\end{eqnarray}
Thus from (\ref{c2mr}), we obtain
\begin{eqnarray}\label{cr2l}
C_{\underbrace{2\cdots 2}_{l}}(N)=\prod_{j=0}^{l-1}(N^{2}+2j).
\end{eqnarray}

By collecting the coefficients of $t_{i_{1}}t_{i_{2}}\cdots t_{i_{k}}$ in (\ref{wm3}) and setting to zero,
we have
\begin{eqnarray}\label{c2}
C_{\underbrace{2\cdots 2}_{l+1}\displaystyle i_{1}\cdots i_{k}}(N)&=&(N^{2}+i_{1}+\cdots +i_{k}+2l)
C_{\underbrace{2\cdots 2}_{l}\displaystyle i_{1}\cdots i_{k}}(N)\nonumber\\
&=&\prod_{j=0}^{l}(N^{2}+i_{1}+\cdots +i_{k}+2j)C_{i_{1}\cdots i_{k}}(N),\,l\in \mathbb{N}.
\end{eqnarray}

Let us take the constraint operator $W^2_0$ in (\ref{wcon}), i.e.,
\begin{eqnarray}\label{cow20}
(\hat{W}_{-2}-\hat{D})Z_{G}=0.
\end{eqnarray}
After a straightforward calculation of the left-hand side of (\ref{cow20}),
we obtain
\begin{eqnarray}\label{w2}
&&\sum_{j_1,j_2=1}^{\infty}(j_1+j_2+2)t_{j_1+j_2+2}(C_{j_1j_2}(N)
+\sum_{i_1=1}^{\infty}C_{j_1j_2i_1}(N)t_{i_1}+\frac{1}{2!}
\sum_{i_1,i_2=1}^{\infty}C_{j_1j_2i_1i_2}(N)t_{i_1}t_{i_2}+\cdots)\nonumber\\
&&+t_{1}^{2}N(1+\sum_{i_1=1}^{\infty}C_{i_1}(N)t_{i_1}+\frac{1}{2!}
\sum_{i_1,i_2=1}^{\infty}C_{i_1i_2}(N)t_{i_1}t_{i_2}+\frac{1}{3!}
\sum_{i_1,i_2,i_3=1}^{\infty}C_{i_1i_2i_3}(N)t_{i_1}t_{i_2}t_{i_3}+\cdots)\nonumber\\
&&+2t_{2}N^{2}(1+\sum_{i_1=1}^{\infty}C_{i_1}(N)t_{i_1}+\frac{1}{2!}
\sum_{i_1,i_2=1}^{\infty}C_{i_1i_2}(N)t_{i_1}t_{i_2}+\frac{1}{3!}
\sum_{i_1,i_2,i_3=1}^{\infty}C_{i_1i_2i_3}(N)t_{i_1}t_{i_2}t_{i_3}+\cdots)\nonumber\\
&&+\sum_{\substack{j_1,j_2=1 \\ j_1+j_2>2}}^{\infty}j_1j_2t_{j_1}t_{j_2}(C_{j_1+j_2-2}(N)
+\sum_{i_1=1}^{\infty}C_{j_1+j_2-2,i_1}(N)t_{i_1}+\frac{1}{2!}
\sum_{i_1,i_2=1}^{\infty}C_{j_1+j_2-2,i_1,i_2}(N)t_{i_1}t_{i_2}\nonumber\\
&&+\cdots)+2\sum_{j_1=1}^{\infty}(j_1+2)t_{j_1+2}N(C_{j_1}(N)+\sum_{i_1=1}^{\infty}C_{j_1i_1}(N)t_{i_1}+\frac{1}{2!}
\sum_{i_1,i_2=1}^{\infty}C_{j_1i_1i_2}(N)t_{i_1}t_{i_2}+\cdots)\nonumber\\
&&-\sum_{j_1=1}^{\infty}j_1t_{j_1}(C_{j_1}(N)+\sum_{i_1=1}^{\infty}C_{j_1i_1}(N)t_{i_1}+\frac{1}{2!}
\sum_{i_1,i_2=1}^{\infty}C_{j_1i_1i_2}(N)t_{i_1}t_{i_2}+\cdots)
=0.
\end{eqnarray}

By collecting the coefficients of $t_1^2$ and setting to zero, we obtain
\begin{eqnarray}\label{c11}
C_{1,1}(N)=N.
\end{eqnarray}
Similarly, for the case of the coefficients of $t_1^l$ with $l$ even,
we have
\begin{eqnarray}\label{w1cij}
C_{\underbrace{1,\cdots, 1}_{l}}(N)=(l-1)NC_{\underbrace{1,\cdots ,1}_{l-2}}(N).
\end{eqnarray}
Substituting (\ref{c11}) into the recursive relation (\ref{w1cij}), we obtain
\begin{eqnarray}
C_{\underbrace{1,\cdots, 1}_{l}}(N)=(l-1)!!N^{\frac{l}{2}},\  for \ l \ even.
\end{eqnarray}

Motivated by the exact $l$-point correlators $C_{\underbrace{1,\cdots, 1}_{l}}(N)$
and $C_{\underbrace{2,\cdots, 2}_{l}}(N)$, we now proceed to derive
the general $l$-point correlators $C_{i_{1}\cdots i_{l}}(N)$.
Let us consider the Virasoro constraints in (\ref{wcon})
\begin{equation}\label{vcon}
W_m^{2}Z_{G}=0.
\end{equation}
The constraint operators $W_m^{2}$ also yield the Witt algebra (\ref{witta})
and null $3$-algebra (\ref{n3witt}).
By means of (\ref{crdw}) and (\ref{dwgr}), we may rewrite (\ref{vcon}) as
\begin{eqnarray}\label{vmrs}
(\hat{W}_{-2})^{m+1}Z_G=
\prod_{j=0}^{m}(\hat{D}-2j)Z_G.
\end{eqnarray}

Let us focus on the coefficients of $t_{i_1}t_{i_2}\cdots t_{i_l}$ 
with $\sum\limits_{j=1}^{l}i_j=2(m+1)$ on the both sides of (\ref{vmrs}).
Note that the form of  $(\hat{W}_{-2})^{m+1}$ appears to become more complicated
very rapidly as one proceeds to higher power.
We may formally express the $(m+1)$-th power of $\hat{W}_{-2}$ as
\begin{eqnarray}\label{hpw2}
(\hat{W}_{-2})^{m+1}&=&\sum_{k,l=1}^{2(m+1)}\sum_{j_1,j_2,\cdots,j_k=0}^{\infty}
\sum_{\substack{i_1+i_2+\cdots+i_l=\rho\\i_1,i_2,\cdots,i_l\geqslant1}}
P^{i_1,i_2,\cdots,i_l}_{j_1,j_2,\cdots,j_k}t_{i_1}\cdots t_{i_l}
\frac{\partial}{\partial t_{j_1}}\cdots \frac{\partial}{\partial t_{j_k}},
\end{eqnarray}
where
$\rho=\sum\limits_{n=1}^k j_n+2(m+1)$ and
$P^{i_1,i_2,\cdots,i_l}_{j_1,j_2,\cdots,j_k}$ are polynomials 
in $i_{\alpha},\,\alpha=1,\cdots,l$ and $j_{\beta},\, \beta=1,\cdots,k$ .

When $j_{\beta}=0$ for $\beta=1,\cdots,k$ in (\ref{hpw2}), the corresponding terms
acting on $Z_{G}$ give the coefficients of $t_{i_1}t_{i_2}\cdots t_{i_l}$ 
with $\sum\limits_{j=1}^{l}i_j=2(m+1)$ on the left-hand side of (\ref{vmrs})
\begin{eqnarray}\label{lhczg}
\sum_{k=1}^{2(m+1)}\sum_{\sigma}
P^{\sigma(i_1),\sigma(i_2),\cdots,\sigma(i_l)}_{\underbrace{0,\cdots, 0}_{k}} N^ke^{Nt_0},
\end{eqnarray}
where $\sigma$ denotes all distinct permutations of $(i_1,i_2,\cdots,i_l)$. 
By means of (\ref{dzs}), the right-hand side of (\ref{vmrs}) becomes
\begin{eqnarray}\label{rhzg}
&&\prod_{j=0}^{m}(\hat{D}-2j)Z_G
=\sum_{s=0}^{\infty}\prod_{j=0}^{m}(s-2j)Z_G^{(s)}\nonumber\\
&&=e^{Nt_0}\sum_{s,l=0}^{\infty}\sum_{\substack{i_1+\cdots +i_l=s\\i_1,\cdots,i_l\geqslant1}}
\frac{1}{l!}\prod_{j=0}^{m}(s-2j)C_{i_{1}\cdots i_{l}}(N)t_{i_1}\cdots t_{i_l}.
\end{eqnarray}
From (\ref{rhzg}), we obtain that the coefficients of $t_{i_1}t_{i_2}\cdots t_{i_l}$ 
with $\sum\limits_{j=1}^l i_j=2(m+1)$ on the right-hand side of (\ref{vmrs}) are
\begin{eqnarray}\label{rhczg}
e^{Nt_0}\sum_{\sigma }\frac{2^{m+1}(m+1)!}{l!}
C_{{\sigma(i_1)},\cdots,{\sigma(i_l)}}(N)=\frac{2^{m+1}(m+1)!\lambda_{(i_1\cdots i_l)}}{l!}e^{Nt_0} C_{{i_1}\cdots{i_l}}(N),
\end{eqnarray}
where we denote by $\lambda_{(i_1\cdots i_l)}$ the
number of  distinct permutations of $(i_1,i_2,\cdots,i_l)$.

By equating (\ref{lhczg}) and (\ref{rhczg}), we obtain 
the $l$-point correlators $C_{i_{1}\cdots i_{l}}(N)$
\begin{eqnarray}\label{reslpc}
C_{{i_1}\cdots{i_l}}(N)=\frac{l!}{2^{m+1}(m+1)!\lambda_{(i_1\cdots i_l)}}\sum_{k=1}^{2(m+1)}
\sum_{\sigma}
P^{\sigma(i_1),\sigma(i_2),\cdots,\sigma(i_l)}_{\underbrace{0,\cdots, 0}_{k}} N^k,
\end{eqnarray}
where $\sum\limits_{j=1}^l i_j$ is even and $m=\frac{1}{2}\sum\limits_{j=1}^l i_j-1$.

When particularized to the 1-point correlators in (\ref{reslpc}),  we have
\begin{eqnarray}
C_{i}(N)=\frac{1}{\sqrt {2^i}(\frac{i}{2})!}\sum_{k=1}^{i}
P^i_{\underbrace{0,\cdots, 0}_{k}} N^k.
\end{eqnarray}
Comparing (\ref{reslpc}) with (\ref{MMC1}) and (\ref{MMC2}), we see that
(\ref{reslpc}) is different from the other two expressions.
Hence (\ref{reslpc}) is a new formula for correlators,
where the operators $(\hat{W}_{-2})^{m+1}$ play an crucial role to determinate 
the polynomials in $i_{\alpha},\,\alpha=1,\cdots,l$ in the correlators.

For clarity of calculation,
let us consider the $m=1$ case in (\ref{reslpc}), i.e., $\sum\limits_{j=1}^l i_j=4$.
From the expression
\begin{eqnarray}
&&(\hat{W}_{-2})^{2}=
\sum_{i_3,i_4=0}^{\infty}\sum_{i_1+i_2=i_3+i_4+4}(i_3+i_4+2)\tilde{t}_{i_1}
\tilde{t}_{i_2}\frac{\partial}{\partial t_{i_3}}\frac{\partial}{\partial t_{i_4}}
+2\sum_{i_1,i_2=0}^{\infty}i_1i_2\tilde{t}_{i_1+i_2+2}\frac{\partial}
{\partial t_{i_1+i_2-2}}\nonumber\\
&&+2\sum_{i_1,i_2,i_3=0}^{\infty}(i_1+i_2-2)\tilde{t}_{i_1}\tilde{t}_{i_2}
\tilde{t}_{i_3}\frac{\partial}{\partial t_{i_1+i_2+i_3-4}}
+2\sum_{i_1,i_2,i_3,i_4=0}^{\infty}\tilde{t}_{i_1}\tilde{t}_{i_2}\tilde{t}_{i_3+i_4+2}
\frac{\partial}{\partial t_{i_1+i_2-2}}\frac{\partial}{\partial t_{i_3}}\frac{\partial}{\partial t_{i_4}}\nonumber\\
&&+\sum_{i_1,i_2,i_3,i_4=0}^{\infty}\tilde{t}_{i_1}\tilde{t}_{i_2}\tilde{t}_{{i}_3}
\tilde{t}_{i_4}\frac{\partial}{\partial t_{i_1+i_2-2}}\frac{\partial}{\partial t_{i_3+i_4-2}}
+\sum_{i_1,i_2,i_3,i_4=0}^{\infty}\tilde{t}_{i_1+i_2+2}\tilde{t}_{i_3+i_4+2}\frac{\partial}{\partial t_{i_1}}
\frac{\partial}{\partial t_{i_2}}\frac{\partial}{\partial t_{i_3}}\frac{\partial}{\partial t_{i_4}}\nonumber\\
&&+4\sum_{i_1,i_2,i_3=0}^{\infty}i_2\tilde{t}_{i_1+i_2+2}\tilde{t}_{i_3}
\frac{\partial}{\partial t_{i_1}}\frac{\partial}{\partial t_{i_2+i_3-2}}
+2\sum_{i_1,i_3,i_4=0}^{\infty}(i_3+i_4+2)\tilde{t}_{i_1+i_3+i_4+4}\frac{\partial}
{\partial t_{i_1}}\frac{\partial}{\partial t_{i_3}}\frac{\partial}{\partial t_{i_4}},
\end{eqnarray}
where $\tilde{t}_{j}=jt_{j}$,
we have
\begin{eqnarray}\label{examp1}
&&P^4_0=8,\quad P^4_{0,0,0}=16,\quad P^{1,3}_{0,0}=6,\quad P^{3,1}_{0,0}=18,\quad P^{2,2}_{0,0}=8,\nonumber\\
&&P^{2,2}_{0,0,0,0}=4,\quad P^{1,1,1,1}_{0,0}=1,\quad P^{1,2,1}_{0}=P^{2,1,1}_{0}=P^{1,1,2}_{0,0,0}=4.
\end{eqnarray}
Substituting (\ref{examp1}) into (\ref{reslpc}), we obtain
\begin{eqnarray}\label{exampc1}
&&C_{4}(N)=\frac{1}{2^2\cdot2!\cdot\lambda_{(4)}}(P^4_0N+P^4_{0,0,0}N^3)=2N^{3}+N,\nonumber\\
&&C_{1,3}(N)=\frac{2!}{2^2\cdot2!\cdot\lambda_{(1,3)}}(P^{1,3}_{0,0}+P^{3,1}_{0,0})N^2=3N^2,\nonumber\\
&&C_{2,2}(N)=\frac{2!}{2^2\cdot2!\cdot \lambda_{(2,2)}}(P^{2,2}_{0,0}N^2+P^{2,2}_{0,0,0,0}N^4)=N^4+2N^2,\nonumber\\
&&C_{1,1,2}(N)=\frac{3!}{2^2\cdot2!\cdot\lambda_{(1,1,2)}}[(P^{1,2,1}_{0}+P^{2,1,1}_{0})N
+P^{1,1,2}_{0,0,0}N^3]=N^3+2N,\nonumber\\
&&C_{1,1,1,1}(N)=\frac{4!}{2^2\cdot2!\cdot \lambda_{(1,1,1,1)}}P^{1,1,1,1}_{0,0}N^2=3N^2,
\end{eqnarray}
where
$\lambda_{(4)}=1$, $\lambda_{(1,3)}=2$, $\lambda_{(2,2)}=1$, $\lambda_{(1,1,2)}=3$ and $\lambda_{(1,1,1,1)}=1$.

%%%%%%%%%%%%%%%%%%%%%%%%%%%%%%%%%%%%%%%%%%%%%%%%%%%%%%%%%%%
\section{Summary}
%%%%%%%%%%%%%%%%%%%%%%%%%%%%%%%%%%%%%%%%%%%%%%%%%%%%%%%%%%%

It is known that the partition function of the Gaussian Hermitian 
matrix model can be obtained by acting on an elementary function 
with exponent of  the operator $\hat W_{-2}$. This operator increases 
the grading in the sense (\ref{increa}). Based on the operators 
$\hat W_{-2}$ and $\hat D$ preserving the grading, we have constructed 
the $W_{1+\infty}$ constraints (\ref{wcon}) for the Gaussian model, where
the constraint operators yield not only the $W_{1+\infty}$ algebra,
but also the closed $W_{1+\infty}$ $n$-algebra. In contrast with the operator 
$\hat{W}_{-2}$, we observed that the operator $\frac{\partial}{\partial t_{2}}$
decreases the grading in the sense (\ref{decre}). Another $W_{1+\infty}$ 
constraints (\ref{nvcowc}) for the Gaussian model have been presented 
in terms of the operators $\frac{\partial}{\partial t_{2}}$ and $\hat D$, 
where the constraint operators also constitute the closed $W_{1+\infty}$ ($n$-)algebras.
When particularized to the Virasoro constraints in (\ref{wcon}) and (\ref{nvcowc}), 
respectively, the corresponding constraint operators give the null 3-algebra.

Based on the Virasoro constraints (\ref{nvcm2}), we have presented the exact 
correlators $C_{{2}\cdots{2}}(N)$ (\ref{cr2l}). However, it appears to be impossible 
to obtain arbitrary correlators from (\ref{nvcm2}). With the help of another Virasoro 
constraints (\ref{vcon}), we have derived a new formula (\ref{reslpc}) for correlators 
in the Gaussian Hermitian matrix model. Our results confirm that the constraint operators 
which lead to the higher algebraic structures  provide new insight into the matrix models.

\section *{Acknowledgments}
We would like to thank the referee for his/her helpful comments.
This work is supported by the National Natural Science Foundation
of China (Nos. 11875194, 11871350 and 11605096).

%%%%%%%%%%%%%%%%%%%%%%%%%%%%%%%%%%%%%%%%%%%%%%%%%%%%%%%%%%%%%%%%%%%%%%%%%%%%%%%

\end{document}